%% file: 20261120_arXiv.tex
\documentclass[reprint,aps,prb,amsmath,amssymb,footinbib,longbibliography]{revtex4-2}
\usepackage{graphicx}
\usepackage{bm}
\usepackage{multirow}
\usepackage[colorlinks=true,linkcolor=blue,citecolor=blue,urlcolor=blue]{hyperref}

\begin{document}

\title{X-ray spectroscopic responses to magnetoelastic symmetry breaking in altermagnets}

\author{N. Sasabe}
\affiliation{Center for Basic Research on Materials (CBRM), National Institute for Materials Science (NIMS), Tsukuba, Ibaraki 305-0047, Japan}
\author{Y. Ishii}
\affiliation{Center for Basic Research on Materials (CBRM), National Institute for Materials Science (NIMS), Tsukuba, Ibaraki 305-0047, Japan}
\author{H. Koizumi}
\affiliation{Department of Applied Physics, University of Tsukuba, Tsukuba, Ibaraki 305-8573, Japan}
\author{Y. Yamasaki}
\affiliation{Center for Basic Research on Materials (CBRM), National Institute for Materials Science (NIMS), Tsukuba, Ibaraki 305-0047, Japan}
\affiliation{International Center for Synchrotron Radiation Innovation Smart, Tohoku University, Sendai 980-8577, Japan}
\affiliation{Center for Emergent Matter Science (CEMS), RIKEN, Wako 351-0198, Japan}

\begin{abstract}
The multipole description of x-ray absorption spectroscopy (XAS), established through the sum rules of x-ray dichroism, provides a framework to describe magnetic and orbital responses in terms of ferroic multipole order parameters.
X-ray magnetic circular dichroism (XMCD) is known to probe spin, orbital, and anisotropic magnetic dipole moments.
Its applications to altermagnets and noncollinear antiferromagnets have revealed that the XMCD response is often governed by
the ferroic component of the photoexcited electronic states.
In this work, we apply the multipole-based analysis of x-ray magnetic linear dichroism (XMLD) to altermagnets and demonstrate that XMLD response in altermagnets can be understood as a manifestation of
magnetoelastic
couplings between magnetic dipole and electric quadrupole moments.
Using symmetry analysis combined with exact diagonalization calculations of $L_{2,3}$-edge XAS, we systematically investigate representative altermagnets, including $\alpha$-MnTe, MnF$_2$, and CrSb.
We show that the ferroic ordering of higher-rank magnetic multipoles, particularly spinful magnetic octupoles,
gives rise to characteristic field-odd XMLD
responses that directly reflect the underlying piezomagnetic response tensors allowed by magnetic point-group symmetry.
Furthermore, we discuss XMCD
responses induced by piezomagnetic effects, in which strain generates magnetic dipole moments.
Our results establish XMLD and XMCD responses as element-specific probes of magnetoelastic multipole order in altermagnets and provide a general symmetry-based pathway to identify hidden ferroic multipoles and strain-controllable spin phenomena beyond conventional ferromagnetism.
\end{abstract}

\maketitle

\section{Introduction}
\subsection{Altermagnets}
Magnetic materials have long been classified into ferromagnets, ferrimagnets, and antiferromagnets based on net magnetization and magnetic ordering \cite{kittel2004issp, blundell2001magnetism, coey2010magnetism, baltz2018antiferromagnetic}.
In conventional \textit{collinear} antiferromagnets, the combination of spatial inversion and time-reversal symmetry preserves spin degeneracy, thereby forbidding momentum-dependent spin splitting \cite{Noda2016_PhysChemChemPhys, Okugawa_AM2018, Ahn2019PRB, Naka2019NatCom, Smejkal2020_SciAdv}.
This classification has been broadened by the theoretical prediction of a new class of magnets termed \textit{altermagnets}: collinear magnets with zero net magnetization yet with spin-split electronic bands \cite{Smejkal2022, Smejkal2022_PRX, Smejkal2023_NatRevMater}.
The spin splitting does not originate from spin-orbit coupling (SOC); rather, it is enforced by magnetic space-group symmetries that act differently on spin-up and spin-down channels in momentum space \cite{HayamiJPSJ2019}.
Thus, these systems break global spin degeneracy while preserving a joint symmetry of crystal operations with time reversal, allowing momentum-dependent spin polarization without net magnetization.

In antiferromagnets, the symmetry of the magnetic multipole order plays a central role in determining observable macroscopic responses.
For example, in systems such as Mn$_3$Sn and the $g$-wave altermagnet MnTe, whose magnetic structures effectively possess magnetic dipole symmetry, phenomena analogous to those in ferromagnets have been observed or theoretically anticipated, including x-ray magnetic circular dichroism (XMCD), the magneto-optical Kerr effect (MOKE), and the anomalous Hall effect (AHE) \cite{Nakatsuji2015, Mn3Sn_2017_Ikhlas_LargeNernst, Mn3Sn_2018_MOKE_higo, MnTe2024Nature}.
In these materials, the broken time-reversal symmetry, combined with appropriate crystal symmetry, enables finite Berry curvature and transverse transport responses despite the absence of net magnetization.
By contrast, $d$- and $g$-wave altermagnets with magnetic octupole symmetry are characterized by spin splitting without net Berry curvature, thereby suppressing the conventional AHE while still permitting unconventional responses \cite{KarubePRL2022, koizumi2023quadrupole, Koizumi_2025}.
In particular, higher-order altermagnetic states, such as $g$-wave and $f$-wave types, are predicted to host transverse spin currents even in the absence of SOC \cite{Rafael2021PRL, Hayami2024PRB, HayamiJLPEA2024}, as well as symmetry-allowed
magnetoelastic coupling that directly reflect the multipolar nature of the underlying spin splitting \cite{BhowalPRX2024, Yuan2024PRL, Belashchenko2025PRL}.
These features highlight a qualitative distinction between dipolar antiferromagnets,
which can mimic ferromagnetic responses, and higher-rank altermagnets,
where transport and magnetoelastic phenomena are governed by multipolar symmetry rather than Berry-curvature-driven mechanisms.

\begin{figure*}[t]
\centering
\includegraphics[width=0.9\textwidth]{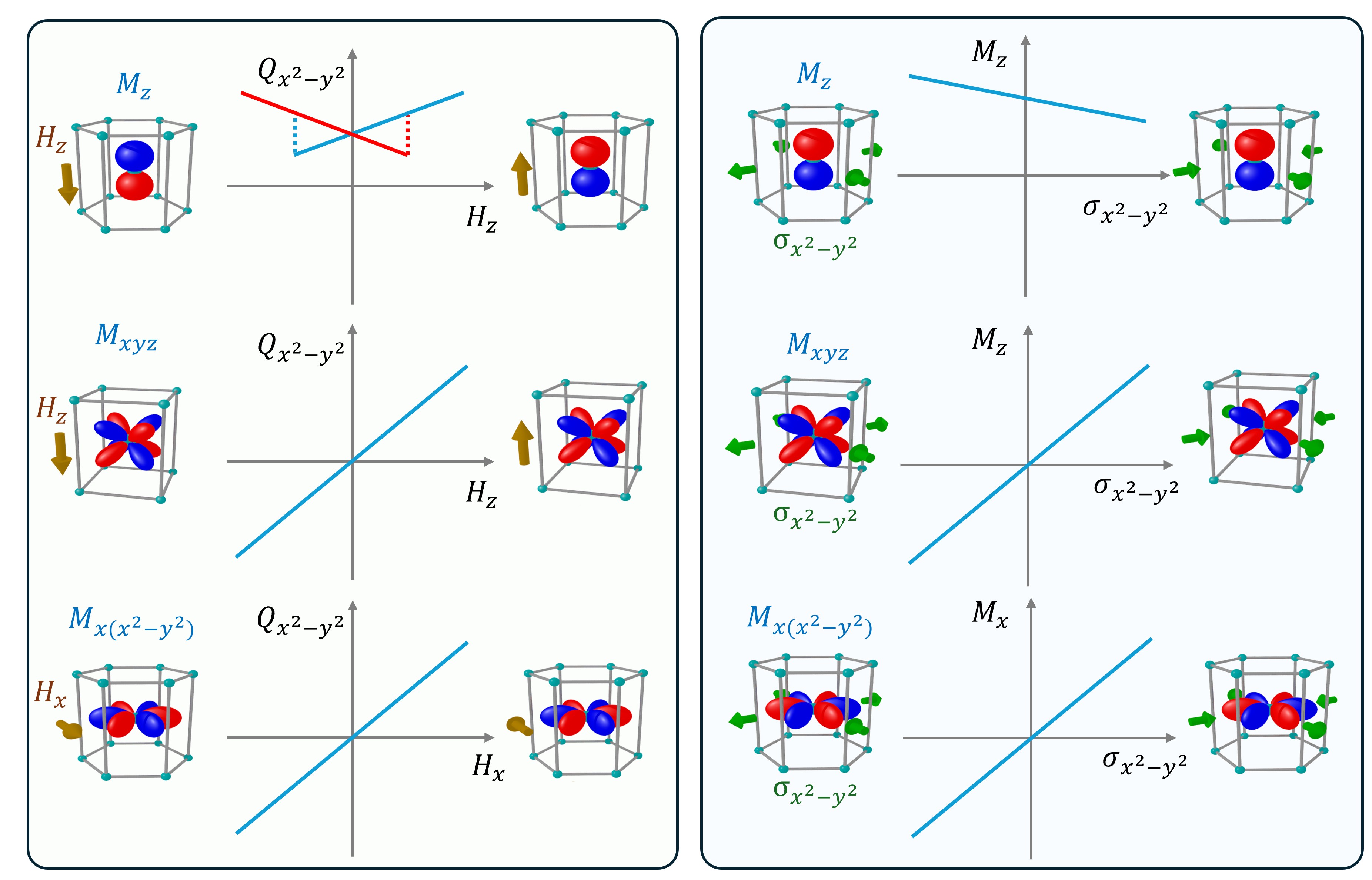}
\caption{
Schematic illustration of multipole-resolved x-ray spectroscopic responses in altermagnetic systems.
Each row corresponds to an altermagnetic symmetry characterized by the lowest-rank symmetry-allowed active magnetic multipoles:
the magnetic dipole $M_z$, the magnetic octupole $M_{xyz}$, or the magnetic octupole $M_{x(x^2-y^2)}$.
Left: Magnetic-field dependence of the electric quadrupole component $Q_{x^2-y^2}$ probed by x-ray magnetic linear dichroism.
Brown arrows indicate the direction of the applied magnetic field.
Right: Strain dependence of the magnetic dipole component probed by x-ray magnetic circular dichroism.
Green arrows indicate the direction of the applied orthorhombic stress $\sigma_{x^2-y^2}$.
The diagrams schematically illustrate how an external magnetic field or strain generates symmetry-allowed multipolar responses characteristic of the corresponding dipolar, $d$-wave, and $g$-wave altermagnetic states.
}
\label{fig:magstr}
\end{figure*}

\subsection{Spinless and spinful multipoles}
From a symmetry viewpoint, each magnetic point group corresponds to a definite pattern of spatial and time-reversal symmetry breaking that can be identified with a leading ferroic magnetic multipole (dipole, quadrupole, octupole, \textit{etc.}) transforming as the same irreducible representation \cite{Mn3Sn_2017_MTSuzuki_Multipole, Hayami2018PRB, SuzukiPRB2019, kusunose2020complete, Yatsushiro2021PRB, hayami2024unified}.
This perspective unifies complex antiferromagnetic orders as manifestations of ferroic multipoles and provides a robust route to predict response tensors purely from symmetry.

Multipole expansions offer a systematic description of charge and magnetic distributions \cite{Kusunose2008JPSJ, Kuramoto2009JPSJ}.
Beyond the conventional electric ($\bm{Q}$) and magnetic ($\bm{M}$) multipoles, the \textit{complete} basis incorporates electric toroidal ($\bm{G}$) and magnetic toroidal ($\bm{T}$) \footnote{The symbol \(T\) used here denotes a magnetic toroidal multipole and should not be confused with the \(T_z\) term in the XMCD sum rules of Thole and Carra~\cite{Thole1992, Carra1993}.
The latter is the \(z\) component of the intra-atomic spin magnetic dipole operator,
or anisotropic magnetic dipole term, and corresponds in the present classification to a spinful magnetic multipole \(M\), not to a magnetic toroidal multipole \(T\).} multipoles and their spinful counterparts are obtained by coupling orbital tensors with spin degrees of freedom~\cite{kusunose2020complete,hayami2024unified}.
Let $\hat{X}_{\ell m}^{(\mathrm{orb})}$ and $\hat{\sigma}_{s,n}$ denote a spinless orbital multipole ($X \in \{Q,M,T,G\}$) and a spin-space basis operator, respectively.
The orbital multipole operators $\hat{X}_{\ell m}^{(\mathrm{orb})}$ are constructed on the basis of spherical harmonics.
Throughout this work, $\hat{X}_{\ell m}^{(\mathrm{orb})}$ follows the standard Condon-Shortley phase convention: $\hat{X}_{\ell m}^{\dagger} = (-1)^m \hat{X}_{\ell,-m}$.
The spinless ($s=0$) and spinful ($s=1$) operators are constructed via angular momentum coupling as
\begin{equation}
  \hat{X}_{\ell m}^{(s,k)}
  = i^{\,s+k}\!\!\sum_{n=-s}^{s}
  C_{\ell+k,\,m-n;\,s n}^{\ell m}\,
  \hat{X}_{\ell+k,\,m-n}^{(\mathrm{orb})}\,
  \hat{\sigma}_{s n},
  \label{eq:spinful}
\end{equation}
where $C_{l_1,m_1;l_2,m_2}^{l,m}$ is the Clebsch-Gordan coefficient.
The parameter $k$ specifies the rank shift of the orbital multipole under spin-orbit coupling,
such that an orbital multipole of rank $\ell+k$ is coupled with a spin tensor of rank $s$ to form a multipole operator of rank $\ell$.
Accordingly, $k$ takes values in the range $k\in\{-s,\ldots,s\}$.
Here, $\hat{\sigma}_{00}$ is the identity matrix $\hat{\sigma}_{00}= \hat{I}$,
while $\hat{\sigma}_{1n}$ ($n=0,\pm1$) are defined from the Pauli matrices:
$\hat{\sigma}_{10}=\hat{\sigma}_z$,
$\hat{\sigma}_{1,\pm1}=\mp(\hat{\sigma}_x\pm i\hat{\sigma}_y)/\sqrt{2}$.
Since spin is an axial vector and is odd under time reversal, the spinful $\hat{X}_{\ell m}^{(1,k)}$ possesses transformation properties opposite in time-reversal parity to the orbital tensor with which it couples.
In $d$-electron systems, for example, a spinful magnetic multipole may combine orbital tensors of different spatial parities, enabling higher-rank magnetic octupoles that are ferroic even in centrosymmetric antiferromagnets \cite{BhowalPRX2024}.

Since spinful multipoles can serve as ferroic order parameters in antiferromagnets, identifying their experimental signatures is a central issue in multipole physics.
X-ray absorption spectroscopy offers a powerful means of experimentally distinguishing spinless from spinful multipoles \cite{YamasakiSTAM2025}.
XMCD and related magneto-optical responses in altermagnets have demonstrated that the anisotropic magnetic dipole (AMD),
corresponding to the spinful magnetic dipole $M_{1m}^{(1,1)}$, serves as a microscopic order parameter in systems with ferroically aligned AMD moments,
giving rise to finite XMCD signals and ferromagnetlike responses even in antiferromagnets \cite{sasabe2023ferroic, Sasabe2025PRB, ishii2026}.
On the other hand, altermagnets characterized by magnetic octupole symmetry, such as $d$-wave MnF$_2$ and $g$-wave CrSb, point to the presence of spinful magnetic octupoles, $M_{3m}^{(1,\pm 1)}$,
which do not couple directly to conventional XMCD \cite{MizumakiPRB2026}.
Such hidden spinful octupolar order is tied to magnetoelastic responses and may be unveiled by detecting the corresponding spinful multipoles through XMCD and XMLD under an external magnetic field or stress.
In this work, we theoretically elucidate how the spinful multipole responses associated with magnetoelastic coupling manifest in XMCD and x-ray magnetic linear dichroism (XMLD) spectra by combining magnetic multipole theory with full-multiplet x-ray absorption spectroscopy (XAS) calculations on representative altermagnets.

The remainder of this paper is organized as follows. In Sec.~II, we formulate the multipolar description of magnetoelastic effects and x-ray absorption spectroscopy. In Sec.~III, we present full-multiplet XAS calculations for representative altermagnets and discuss their XMLD and XMCD responses. Finally, Sec.~IV summarizes the main conclusions.

\section{Multipoles in magnetoelasticity and x-ray absorption spectroscopies}

\subsection{Magnetoelastic coupling}
The magnetoelastic couplings describe the fundamental coupling between magnetic and elastic degrees of freedom,
encompassing phenomena in which magnetic order and lattice deformation mutually influence each other \cite{Dzialoshinskii1958, Dzialoshinskii1958Piezomagnetism, MORIYA195973, Dzyaloshinski1960OnTM, BorovikRomanov1960JETP, Borovik-romanov01011994, AoyamaPRM2024}.
From a phenomenological viewpoint, this coupling manifests itself in two distinct responses: the direct piezomagnetic effect
and its inverse effect.
The total strain response $\varepsilon_{jk}$ of a magnetic system under an external magnetic field $\mathbf{H}$ can be systematically expressed as
\begin{equation}
\varepsilon_{jk}=\Lambda_{jk;i} H_i+N_{jk;mn} H_m H_n,
\end{equation}
where the first and second terms represent the linear inverse-piezomagnetic and quadratic magnetostrictive contributions, respectively.
$\Lambda_{jk;i}$ is the third-rank tensor describing the
inverse piezomagnetic coupling between the magnetic field and strain.
Since the strain is even and the magnetic field is odd under time reversal,
a nonzero $\Lambda_{jk;i}$ requires broken time-reversal symmetry and therefore magnetic order.
In contrast,
the quadratic term, magnetostriction, is a universal feature of magnetic materials and is characterized by a rank-4 tensor $N_{jk;mn}$.
Since it depends on even powers of the magnetic field, the induced strain is invariant under field reversal.
The corresponding direct piezomagnetic effect
governed by the tensor $\Lambda_{i;jk}$, namely, the induction of magnetization $m_i$ by external stress $\sigma_{jk}$, is expressed as
\begin{equation}
m_i = \Lambda_{i;jk} \sigma_{jk}.
\end{equation}
Unlike the piezoelectric effect, which requires broken inversion symmetry, the piezomagnetic effect is symmetry allowed even in certain centrosymmetric crystals, provided that time-reversal symmetry is broken by magnetic ordering \cite{Andratskii1967_JETP, Prokhorov1969_JETPLett, Baruchel1988_JPhysColloq}.

From a microscopic perspective, the inverse piezomagnetic effect can be interpreted as a magnetic-field-induced modification of electric quadrupole moments $Q_{2m}$.
These induced quadrupoles couple to the lattice via electron-lattice interactions, giving rise to a macroscopic strain response \cite{Ogawa2025JPSJ}.
Since lattice strain and stress transform as rank-2 symmetric tensors, they can naturally be identified with electric quadrupole degrees of freedom.
This observation provides a direct link between magnetoelastic responses and magnetic multipole order.
Symmetry analysis shows that the magnetoelastic coupling is allowed in systems hosting ordered magnetic dipoles $M_{1m}$, magnetic toroidal quadrupoles $T_{2m}$, or magnetic octupoles $M_{3m}$ \cite{Huyen2025PRB, Hayami2018PRB, Huyen2019PRB}.
Although magnetostriction may also occur in antiferromagnets, it can be distinguished from the piezomagnetic effect by its even parity with respect to magnetic-field reversal \cite{Callen1965, Callen1968, Meng2024NatCommun, Tomikawa2024PRBL}.

\subsubsection*{Magnetoelastic coupling in altermagnets}
In this work, we focus on altermagnets and analyze their magnetoelastic responses within the framework of magnetic multipole theory.
Since our primary interest is the spectroscopic response in XAS, we restrict the discussion to atomic magnetic multipoles rather than augmented (extended) multipoles.
In other words, we focus only on those components of the extended multipoles that can be mapped onto the atomic picture and are spectroscopically observable, namely, the electronic orders of magnetic dipoles, electric quadrupoles, and magnetic octupoles.

The emergence of piezomagnetic and inverse piezomagnetic effects associated with various magnetic multipole orders can be systematically formulated using the spherical harmonics representation \cite{hayami2024unified}.
Within this framework, the coupling between magnetic multipoles, external fields, and strain is governed by angular momentum selection rules encoded in Clebsch-Gordan coefficients:
\begin{align}
Q_{2m}&=\sum_{n,m^\prime}\left(\alpha C^{2m}_{1m^\prime;1n} M_{1m^\prime}+\beta C^{2m}_{3m^\prime;1n} M_{3m^\prime}\right)H_{1n},\\
M_{1m}&=\sum_{n,m^\prime}\left(\gamma C^{1m}_{1m^\prime;2n} M_{1m^\prime}+\delta C^{1m}_{3m^\prime;2n} M_{3m^\prime}\right)\sigma_{2n}.
\end{align}
Here, $Q_{2m}$, $M_{1m}$, and $M_{3m}$ denote the electric quadrupole, magnetic dipole, and magnetic octupole moments, respectively.
The external magnetic field $H_{1n}$ and the
external mechanical stress field
$\sigma_{2n}$ are expressed in rank-1 and rank-2 spherical tensors, respectively.
The Clebsch-Gordan coefficients $C^{lm}_{l_1 m_1;l_2 m_2}$ impose strict symmetry constraints on the allowed couplings.
Importantly, a given magnetoelastic response is symmetry allowed only when the corresponding Clebsch-Gordan coefficient is nonzero.
This condition directly connects macroscopic piezomagnetic and inverse piezomagnetic responses to the underlying ferroic magnetic multipole order and the angular momentum structure of the electronic degrees of freedom.

The physical implications of Eqs.~(4) and (5) are schematically summarized in Fig.~1, and their relation to Eqs.~(2) and (3) is discussed as follows.
The figure visualizes how the angular momentum selection rules encoded in the Clebsch-Gordan coefficients determine the parity and linearity of magnetoelastic responses for altermagnetic states belonging to different magnetic point groups.
We first consider Eq.~(4), which corresponds to the inverse piezomagnetic
effect expressed in Eq.~(2), namely, the lattice deformation $\epsilon_{jk}$ induced by an external magnetic field $H_{1n}$ via the electric quadrupole $Q_{2m}$.
When a magnetic dipole $M_{1m}$ is the lowest-rank active multipole of the altermagnetic state, the $Q_{x^2-y^2}$-type inverse-piezomagnetic response to $H_z$ considered here originates from the higher-rank magnetic octupole $M_{3m^\prime}$ ($m^\prime=\pm2$).
As illustrated in the upper-left panel of Fig.~1, this octupole coupling gives rise to an odd (asymmetric) dependence of $Q_{x^2-y^2}$ on $H_z$ in the low-field region, where the ferroic AMD order ($M_{1m}^{(1,1)}$) remains unchanged.
Under sufficiently strong magnetic fields, however, the N\'eel vector can be reoriented via the Dzyaloshinskii-Moriya interaction, as observed in MnTe and FeS \cite{FeS2025NatMat,MnTe2024Nature}.
This reorientation reverses the ferroic AMD order together with the associated octupolar component responsible for the magnetoelastic coupling.
As a result, $Q_{x^2-y^2}$ exhibits an even (symmetric) field dependence in the high-field region, characteristic of conventional magnetostriction in magnets.

In contrast, when the primary order parameter is a magnetic octupole $M_{3m^\prime}$, the relevant coupling is determined by $C^{2m}_{3m^\prime;1n}$.
Since the octupole does not transform as a dipole, its response to the magnetic field is fundamentally different.
The induced quadrupole arises linearly from the product $M_{3m^\prime}H_{1n}$ and changes sign under the reversal of the field or the magnetic octupole domain, as well as the N\'eel vector.
As shown in the middle and lower panels of the left column, this produces an odd (antisymmetric) linear dependence of  $Q_{x^2-y^2}$ on $H_z$ or $H_x$,
which serves as a signature of octupolar ($d$-wave or $g$-wave) altermagnetic order.
Thus, the parity of the inverse piezomagnetic response directly reflects whether the underlying ferroic order is dipolar or octupolar.

We next discuss Eq.~(5), which describes the induction of a magnetic dipole $M_{1m}$ by symmetric strain $\sigma_{2n}$.
This equation corresponds to the piezomagnetic effect given in Eq.~(3).
When a magnetic dipole is the lowest-rank active multipole of the altermagnetic state, the coupling between $\sigma_{x^2-y^2}$ and $M_{3m^\prime}$ ($m^\prime=\pm2$) induces $M_z$, indicating that the magnetic dipole moment is modified by strain.
However, when a magnetic octupole $M_{3m'}$ is present, the term governed by $C^{1m}_{3m';2n}$ leads to a linear conversion of octupolar order into a magnetic dipole under strain, where symmetric strain $\sigma_{x^2-y^2}$ generates a finite magnetic dipole component ($M_z$ or $M_x$)
proportional to the preexisting octupole.
The induced magnetic dipole reverses sign when the strain or the octupole domain is reversed, reflecting the linear and symmetry-controlled nature of the coupling.
In addition, weak ferromagnetic contributions arising from strain-induced Dzyaloshinskii-Moriya interactions may also appear in magnetization measurements.

Importantly, Fig.~1 demonstrates that the even or odd character of the response functions is not material specific but is dictated by the angular momentum addition rules expressed in Eqs.~(4) and (5).
The Clebsch-Gordan coefficients act as symmetry filters that determine which multipole combinations are allowed.
Therefore, measurements of field- or strain-induced quadrupole and magnetization responses provide direct experimental access to the underlying ferroic multipole order and its tensorial structure in altermagnets.

\subsection{X-ray absorption spectrum}
XAS is a powerful probe of electronic structure, providing element-specific and orbital-selective information on unoccupied electronic states.
In XMCD, the difference between left- and right-circularly polarized absorption isolates rank-1 magnetic multipoles.
The sum rules of XMCD have long provided a quantitative framework for extracting fundamental magnetic quantities, such as the spin angular momentum $S_z$, orbital angular momentum $L_z$, and the AMD $T_z$ terms from core-level x-ray absorption spectra \cite{Thole1992, Carra1993, TanakaJo1996JPSJ, vanDerLaan2014}.
Alongside XMCD, spectroscopic probing of various higher-rank multipolar moments has also been firmly established over the past decades through XMLD and generalized sum rules \cite{vanDerLaan1994, vanVeenendaal1996, Lovesey2005}.

Building upon these pioneering works, the electric dipole ($E1$) transition process in core-level XAS has recently been formulated within a complete and orthonormal electronic multipole basis \cite{YamasakiSTAM2025}. This basis is the same as that used in modern condensed matter physics to classify order parameters and macroscopic responses, such as the anomalous Hall effect and magnetoelastic responses \cite{Yatsushiro2021PRB, Hayami2018PRB, hayami2024unified}. Consequently, the same symmetry tables can be applied directly to both the physical properties and the spectroscopic observables, and group-theoretical analysis based on the symmetry of the multipoles provides a practical framework for using XMCD and XMLD to identify multipolar orders, including those under external fields or structural distortions.
Recent studies have revealed that in altermagnets and noncollinear antiferromagnets, such as MnTe \cite{Hariki2024PRL, DalDin2024, YamamotoPRAp2025} and Mn$_3$Sn \cite{XMCD_Kimata, sakamoto2021observation, sakamoto2024bulk}, the XMCD response is dominated not by conventional spin magnetization but by the AMD term  \cite{Yamasaki2020, Sasabe_PRL_Mn3Sn, sasabe2023ferroic, Sasabe2025PRB, ishii2026}, corresponding to $M_{1m}^{(1,1)}$.
The ferroic ordering of AMD moments gives rise to finite XMCD signals and can induce ferromagnetlike anomalous Hall effects despite vanishing net magnetization \cite{hayami2021essential, OhgataPRB2025}.

While XMCD isolates these odd-rank magnetic multipoles, complementary information on even-rank electric quadrupole moments is obtained from XMLD. This probe is defined as the difference in the XAS measured with orthogonal linear polarizations and can be classified into out-of-plane and in-plane components, depending on the polarization geometry \cite{YamasakiSTAM2025}.
The out-of-plane XMLD
reflects the difference between polarizations parallel and perpendicular to a reference axis and primarily probes the spinless electric quadrupole $Q_{20}^{\rm (orb)} \propto (3z^2 - r^2)$, revealing orbital anisotropy and charge redistribution.
In contrast, the in-plane XMLD
measures the anisotropy within the basal plane and is governed by
$Q_{22}^{\rm (orb)} + Q_{2-2}^{\rm (orb)} \propto (x^2 - y^2)$.
Beyond these spinless contributions, XMLD also contains spinful components arising from magnetic dipole-dipole and dipole-octupole couplings, commonly denoted as $P$ and $R$ terms \cite{CARRA1993PhyB, van1999magnetic, YamasakiSTAM2025}.
These higher-order multipoles encode anisotropic spin-quadrupole interactions and provide direct spectroscopic access to magnetocrystalline anisotropy and ferroic higher-rank magnetic order.

Since the x-ray absorption process is described by a one-electron transition operator connecting the core and valence shells,
its selection rules are governed by the symmetry of the ground state and the transition operator.
Within the complete multipole basis, only those ferroically ordered ground-state multipoles belonging to the symmetry channels contained in the transition operator can contribute to the spectroscopic response.
Each spectral feature, however, observed in XAS corresponds to a transition into a final-state multiplet,
whose energy and spectral line shape are governed by the excited-state electronic structure,
including strong core-level spin-orbit coupling, core-hole interactions, crystal-field splitting, and multiplet effects.
Consequently, although the transition intensity is determined by the transition matrix element connecting the ground and final states,
the detailed spectral line shape generally cannot be attributed uniquely to an individual ground-state multipole.
Rather, the existence of a dichroic response is determined by whether the corresponding ground-state multipole belongs to a symmetry channel selected by the transition operator.
Therefore, the observed dichroic response and the underlying ground-state multipole share the same symmetry channel.
By contrast, the energy-integrated XMCD and XMLD intensities obey the corresponding sum rules and quantitatively probe the symmetry-allowed ferroically ordered ground-state multipole components.

\section{Magnetoelastic response on x-ray absorption spectra in altermagnets}

To clarify the electronic structure and the magnetoelastic coupling in representative altermagnets, $\alpha$-MnTe, MnF$_2$, and CrSb, we calculate the electronic ground-state and dipole transitions in multielectron systems, including full-multiplet effects.
The electronic ground-state calculations and XAS simulations were performed using the EDRIXS Python modules \cite{EDRIXS}.
The atomic model adopts the crystal-field parameters calculated from the crystal structure obtained from the CIF files in the materials project \cite{MnTecif, MnF2cif, CrSbcif} and an effective magnetic field arising from the antiferromagnetic interaction that realizes the altermagnetic state.
The effective exchange field was set to 5 meV throughout the calculations; the dependence on this parameter is examined in Appendix.
The full multiplet effects were taken into account based on atomic calculations within the Hartree-Fock-Slater (HFS) method, where the spin-orbit coupling constants and Slater integrals were obtained from self-consistent HFS calculations \cite{cowan1981theory}. The Slater integrals were reduced to 78\% of their atomic values to account for screening effects in the solid state.
For the pressure response, the lattice strain induced by uniaxial pressure was estimated using the stiffness tensor obtained from the Materials Project database, and the corresponding change in the crystal field was evaluated.
The material-specific crystal and magnetic symmetries relevant to these calculations are summarized in Table~I. In particular, the local site symmetries determine the crystal-field environments used in the atomic model, whereas the magnetic point groups constrain the symmetry-allowed multipoles. The multipole classifications adopted in the following discussion are based on the group-theoretical analysis of Yatsushiro \textit{et al.}~\cite{Yatsushiro2021PRB}. Building on these established classifications, we examine how the symmetry-allowed multipoles manifest themselves in the XMCD and XMLD responses through magnetoelastic coupling.

\begin{table*}[t]
\caption{Crystal structures, crystallographic point groups (CPGs), magnetic point groups (MPGs), applied uniaxial stress directions, and the resulting magnetic point groups under uniaxial stress (MPG$_{\rm strain}$) for MnF$_2$, MnTe, and CrSb.
The lowest-rank symmetry-allowed active multipole for each magnetic point group is indicated in parentheses. The classification of the active multipoles, including the magnetic dipole $M_z$ and the magnetic octupoles $M_{xyz}$ and $M_{3a}$, follows the symmetry analysis of Yatsushiro \textit{et al}.~\cite{Yatsushiro2021PRB} In the reduced $m'm'm$ symmetry, the ordinary mirror plane is perpendicular to the [001] axis for MnF$_2$, to the [0001] axis for MnTe, and to the $[\bar{1}100]$ axis for CrSb, thereby allowing a magnetic dipole along the corresponding direction through the piezomagnetic effect.}
\vspace{0.8\baselineskip}
\centering
\setlength{\tabcolsep}{6pt}
\renewcommand{\arraystretch}{1.25}

\begin{tabular}{ccccccc}
\hline
Compound & Crystal structure & CPG & N\'eel vector & MPG & Stress & MPG$_{\rm strain}$ \\
\hline
MnF$_2$ \cite{Yamani2010CJP}
& Rutile
& $D_{4h}$
& $[001]$
& $4'/mm'm$ ($M_{xyz}$)
& $[110]$
& $m'm'm$ ($M_z$) \\
MnTe \cite{kunitomi1964}
& \multirow{2}{*}{NiAs-type}
& \multirow{2}{*}{$D_{6h}$}
& $[\bar1100]$
& $m'm'm$ ($M_z$)
& $[11\bar20]$
& $m'm'm$ ($M_z$) \\
CrSb \cite{Yuan2020Dalton}
&
&
& $[0001]$
& $6'/m'mm'$ ($M_{3a}$)
& $[11\bar20]$
& $m'mm'$ ($M_y$) \\
\hline
\end{tabular}
\end{table*}

\begin{figure*}[t]
    \centering
    \includegraphics[width=0.9\textwidth]{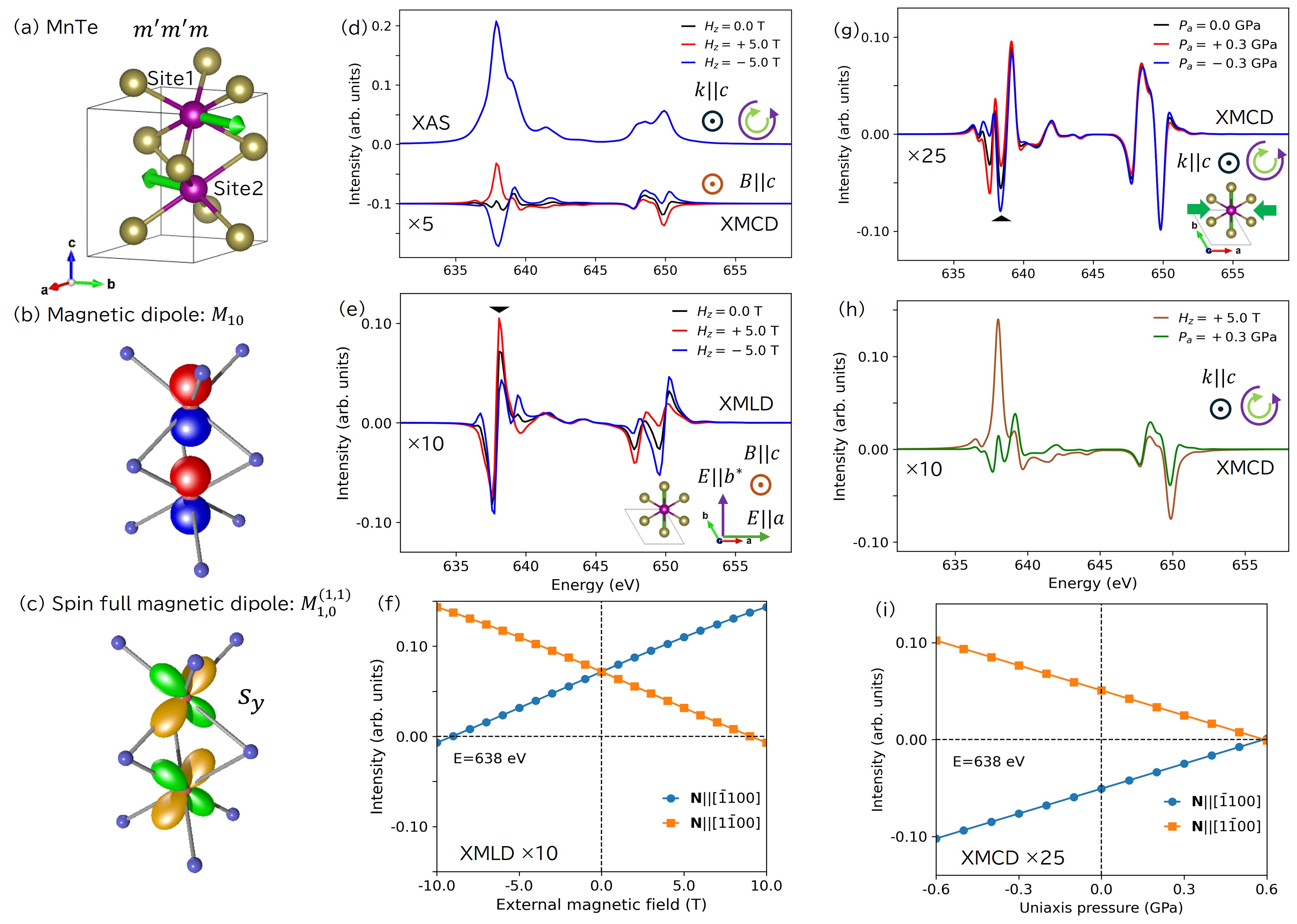}
    \caption{
        Magnetic multipole structure and x-ray absorption spectra in $\alpha$-MnTe.
        (a) Crystal and magnetic structure of $\alpha$-MnTe with the NiAs-type lattice.
        The in-plane N\'eel vector is oriented along the $[1\bar{1}00]$ direction, described by the magnetic point group $m'm'm$.
        The green arrows represent the spin direction.
        This symmetry allows (b) a magnetic dipole $M_{10}$ along the $c$ axis and (c) a spinful magnetic dipole $M_{10}^{(1,1)}$, whose microscopic distribution originates from the $yz$-type quadrupole moment associated with the $s_y$ spin component, where $x \parallel a$, $y \parallel b^{*}$, and $z \parallel c$.
        In the spherical-harmonic representation, red and blue colors denote the sign of the magnetic monopole charge density defined by $\mathbf{m}\cdot\mathbf{r}$, while green and yellow indicate the sign of the spin operator.
        (d) X-ray absorption spectrum (XAS) and x-ray magnetic circular dichroism (XMCD) spectra around the Mn $L$ edge for magnetic fields applied parallel and antiparallel to the $c$ axis.
        (e) X-ray magnetic linear dichroism (XMLD) spectra measured between orthogonal linear polarizations $E\parallel a$ and $E\parallel b$ under opposite magnetic fields.
        (f) Magnetic-field dependence of the XMLD intensity at $E = 638$~eV.
        (g) XMCD spectra under uniaxial stress ($P_{a}=\pm0.3$~GPa) applied perpendicular to the mirror plane.
        (h) Comparison between magnetic-field-induced and uniaxial-stress-induced XMCD spectra.
        (i) Uniaxial-stress dependence of the XMCD intensity at $E = 638$~eV.
        A Lorentzian broadening of 0.42 eV (FWHM) was used for $\alpha$-MnTe. The sign and qualitative line shape of the XMCD/XMLD spectra are robust against reasonable variations of the broadening parameter ($\pm 0.1$ eV).
        }
    \label{fig:xmld}
\end{figure*}

\subsection{$g$-wave altermagnet with magnetic dipole symmetry}

$\alpha$-MnTe crystallizes in the NiAs-type structure, belonging to the hexagonal space group $P6_3/mmc$ (No.~194) \cite{GreenwaldMnTe1953, KomatsubaraJPSJ1963, kunitomi1964}.
Below the N\'eel temperature ($T_N \sim 310$~K \cite{NeelTmpMnTe}), MnTe exhibits a collinear antiferromagnetic order in which the Mn moments are aligned ferromagnetically within each $ab$ plane and coupled antiferromagnetically between adjacent planes along the $c$ axis, as shown in Fig.~2(a).
The N\'eel vector lies in the $ab$ plane, typically along the $[1\bar{1}00]$ direction, lowering the symmetry to the magnetic point group $m'm'm$.
This magnetic point-group symmetry allows for a rank-1 magnetic dipole response $M_{10}$ along the $c$ direction [Fig.~2(b)], even though the net magnetization remains zero.
This symmetry argument naturally explains the appearance of XMCD signals in $\alpha$-MnTe despite its antiferromagnetic ground state.
Microscopically, the XMCD signal originates from the spinful magnetic dipole operator $M_{10}^{(1,1)}$ \cite{Sasabe2025PRB, Yatsushiro2021PRB}, whose spatial distribution corresponds to the $s_y$ spin component coupled to the $yz$-type orbital character,
as illustrated in Fig.~2(c).
Such spin-quadrupole coupling manifests as a spin splitting associated with $s_y$ in the band dispersion \cite{Mazin2023PRB, Osumi2024PRB, Liu2024PRL}. In the parent hexagonal $D_{6h}$ symmetry, this magnetic dipole ($M_{z}$) transforms as the $A_{2g}$ irreducible representation.

Figure~2(d) shows the calculated XAS and XMCD spectra around the Mn $L$ edge for incident x-rays parallel to the $z$ axis ([0001]), obtained by exact diagonalization for the antiferromagnetic state of $\alpha$-MnTe.
Even in the absence of an external magnetic field, a finite XMCD signal appears due to the AMD term, which reflects the symmetry of the antiferromagnetic state.
When a magnetic field is applied along the $[0001]$ direction, an additional XMCD component emerges from field-induced spin polarization.
This contribution originates from a canting of the antiferromagnetic structure and reverses its sign upon reversing the magnetic field.
In contrast, the XMCD component associated with the AMD term is governed by the orientation of the N\'eel vector and therefore does not reverse in the same manner.
Under sufficiently strong magnetic fields, the N\'eel vector can be reoriented via the Dzyaloshinskii-Moriya interaction, leading to a reversal of the AMD-derived XMCD signal.
This mechanism accounts for the reversal of the XMCD response observed in $\alpha$-MnTe \cite{Hariki2024PRL, Sasabe2025PRB}.

When the N\'eel vector is aligned along $[1\bar{1}00]$, the sixfold rotational symmetry of the crystal is reduced,
giving rise to XMLD, a difference in XAS with polarization $E||x$ ([11$\bar{2}$0]) and $E||y$ ([1$\bar{1}$00]) [Fig.~2(e)].
The application of a magnetic field modifies the electronic states and induces a change in the XMLD intensity.
The in-plane XMLD response at zero magnetic field originates from an electronic anisotropy with $E_{2g}$ symmetry in the parent hexagonal structure.
The out-of-plane magnetic-field component $H_z$ transforms as the $A_{2g}$ irreducible representation.
Since
$E_{2g}\otimes A_{2g}=E_{2g}$,
the coupling to $H_z$ preserves the $E_{2g}$ symmetry of the response, allowing the magnetic field to modulate the corresponding XMLD intensity without changing its symmetry character.
In the low-field regime, where the N\'eel vector remains unchanged, the XMLD response is antisymmetric with respect to the magnetic field, reflecting the linear coupling between the field-induced spin canting and the electronic anisotropy.
Figure~2(f) presents the magnetic-field dependence of the XMLD intensity at $E = 638$~eV.
In the low-field region, the XMLD exhibits an antisymmetric dependence on the magnetic field, consistent with field-induced modifications through the linear inverse piezomagnetic effect.

Figure~2(g) shows the change in the XMCD ($\mathbf{k}||[0001]$) spectrum induced by uniaxial stress applied perpendicular to the mirror plane.
The applied strain linearly couples to the underlying magnetic multipole order, thereby modulating the XMCD component.
In the absence of stress, the XMCD response for $\mathbf{k}\parallel[0001]$ originates from the $M_z$-type AMD component $M_{1,0}^{(1,1)}$, which transforms as $A_{2g}$ in the parent hexagonal symmetry.
The magnetic structure also hosts the magnetic octupole $M_{z}^{\beta} (M_{3,\pm2})$, which belongs to the $E_{2g}$ irreducible representation in the parent hexagonal symmetry.
The in-plane uniaxial stress transforms as $E_{2g}$, with components $(2\sigma_{xy},\sigma_{x^2-y^2})$, and microscopically modifies the orbital distribution through the distortion of the ligand field.
The coupling between the $E_{2g}$ magnetic octupole and the $E_{2g}$ stress contains an $A_{2g}$ magnetic component according to
$E_{2g}\otimes E_{2g}=A_{1g}\oplus A_{2g}\oplus E_{2g}$.
This stress-induced $A_{2g}$ component has the same symmetry as the $M_z$-type AMD channel probed by XMCD and therefore modulates the XMCD intensity.
Figure~2(h) compares the XMCD spectra induced by a magnetic field and uniaxial stress.
Since the microscopic origins are distinct, $i.e.$, field-induced spin canting in the former and strain-induced AMD in the latter, the spectral line shapes are different.
Figure~2(i) presents the uniaxial-stress dependence of the XMCD intensity at $E = 638$~eV for the two N\'eel-vector orientations, $\mathbf{N}\parallel[\bar{1}100]$ and $[1\bar{1}00]$.
This leading-order linear and domain-sensitive behavior indicates that the XMCD response is symmetry allowed to couple to the N\'eel vector through the piezomagnetic effect.
The piezomagnetic effect observed in XMCD originates from the AMD term in the excited state,
whereas in magnetization measurements, it is expected to manifest as weak ferromagnetism arising from a strain-induced Dzyaloshinskii-Moriya interaction.

\begin{figure*}[t]
    \centering
    \includegraphics[width=0.9\textwidth]{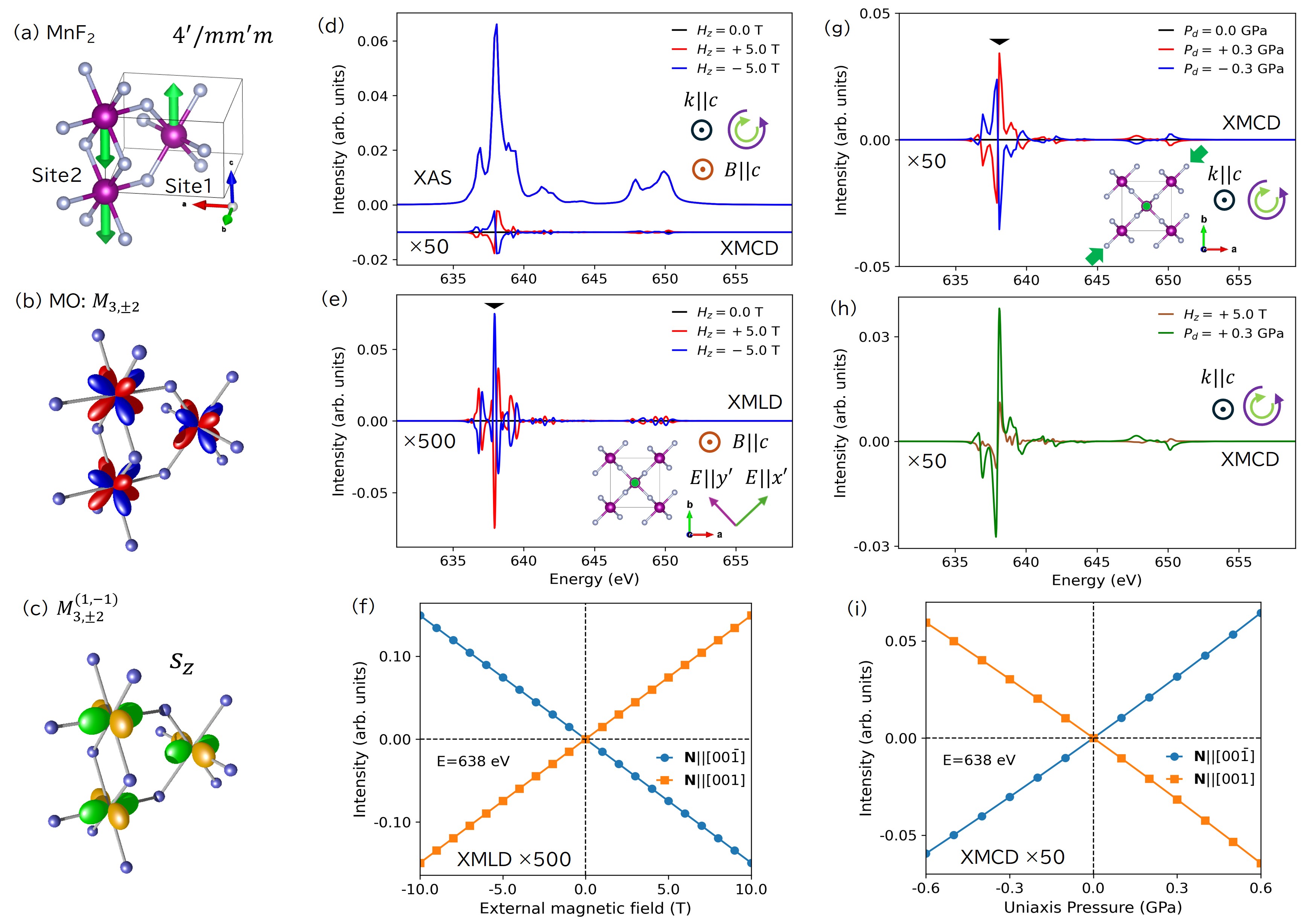}
    \caption{
    Magnetic multipole structure and x-ray absorption spectra in MnF$_2$.
    (a) Crystal and magnetic structure of MnF$_2$ with the rutile structure.
    The antiferromagnetic spins are aligned along the $c$ axis, reducing the symmetry to the magnetic point group $4'/mm'm$.
    This symmetry allows
    (b) the $xyz$-type magnetic octupole moments ($M_{3,\pm2}$), and (c) their spinful counterpart of the $xy$-type quadrupole associated with the $s_z$ spin component, $M_{3,\pm2}^{(1,-1)}$, where $x \parallel a$, $y \parallel b$, and $z \parallel c$.
    (d) X-ray absorption spectrum (XAS) and x-ray magnetic circular dichroism (XMCD) spectra around the Mn $L$ edge for magnetic fields applied parallel and antiparallel to the $c$ axis.
    (e) X-ray magnetic linear dichroism (XMLD) spectra measured between orthogonal linear polarizations ($E\parallel x'$ and $E\parallel y'$) under opposite magnetic fields, where $x'\parallel[110]$ and $y'\parallel[\bar{1}10]$.
    (f) Magnetic-field dependence of the XMLD intensity at $E = 638$~eV.
    The XMLD signal exhibits a linear dependence on the applied magnetic field
    and reverses sign depending on the N\'eel-vector orientation
    ($\mathbf{N}\parallel[001]$ and $[00\bar{1}]$).
    (g) XMCD spectra under uniaxial pressure ($P_{d}=\pm0.3$~GPa).
    (h) Comparison between magnetic-field-induced and uniaxial-pressure-induced XMCD spectra.
    (i) Uniaxial-pressure dependence of the XMCD intensity at $E = 638$~eV.
    A Lorentzian broadening of 0.17 eV (FWHM) was used for MnF$_2$ \cite{MnF2XAS}. The sign and qualitative line shape of the XMCD/XMLD spectra are robust against reasonable variations of the broadening parameter ($\pm 0.1$ eV).
    }
    \label{fig:MnF2}
\end{figure*}

\subsection{$d$-wave altermagnet}
MnF$_2$ adopts the rutile-type crystal structure belonging to the tetragonal space group $P4_2/mnm$ (No.~136) \cite{Griffel1950JACS, Shull1951PR, Yamani2010CJP}.
Below its N\'eel temperature ($T_N \sim 67$~K \cite{NeelTmpMnF2}), the Mn spins order collinearly along the $c$ axis in an antiferromagnetic configuration ($\mathbf{N} \parallel [001]$), resulting in a magnetic symmetry described by the point group $4'/mm'm$ [Fig.~3(a)] \cite{BhowalPRX2024}.
This magnetic structure places MnF$_2$ in the class of $d$-wave altermagnets, where the staggered spin configuration gives rise to a $d_{xy}$-like momentum-dependent spin splitting in the $s_z$ component \cite{Smejkal2022, Smejkal2022_PRX, Smejkal2023_NatRevMater}.
The magnetic point group $4'/mm'm$ strictly forbids rank-1 magnetic dipole responses, ensuring that no net magnetization or dipolar XMCD signal appears in the absence of symmetry breaking~\footnote{Notably, for the $d_{xy}$-like momentum-dependent spin splitting with
$\mathbf{N}\parallel[100]$, the magnetic point group consists of mirror
planes normal to $[100]$ and $[001]$ combined with time reversal, and an
ordinary mirror plane normal to $[010]$, i.e.,
$m'_{[100]}m'_{[001]}m_{[010]}$, which allows XMCD along the $b$ axis~\cite{sasabe2023ferroic}.}.
On the other hand, the rank-3 magnetic octupole components $M_{3,\pm2}$, corresponding to $M_{xyz}$ or $M_{(x^2-y^2)z}$-type distributions \cite{Yatsushiro2021PRB, BhowalPRX2024}, are permitted in the $d$-wave rutile structure [Fig.~3(b)].
Since MnF$_2$ consists of two antiferromagnetically aligned sublattices, the spinless octupole $M_{3,\pm2}^{(0,0)}$ does not form a ferroic order.
Instead, the relevant ferroic order parameter is the spinful magnetic octupole $M_{3,\pm2}^{(1,-1)}$, which combines the $s_z$ spin component with a $d_{xy}$-type orbital distribution [Fig.~3(c)].
In the parent tetragonal $D_{4h}$ symmetry, this magnetic octupole ($M_{xyz}$) transforms as the $B_{1g}$ irreducible representation.
This spin-quadrupole entangled octupole provides the microscopic origin of magneto-optical responses in MnF$_2$ despite its zero net magnetization.

Figure~3(d) shows the calculated XAS and XMCD spectra around the Mn $L$ edge for incident x-rays parallel to the $c$ axis.
Consistent with symmetry, no intrinsic dipolar XMCD signal appears at zero magnetic field.
When a magnetic field is applied along the $c$ direction, a finite XMCD signal is induced through field-driven modulation of the spin component.
This field-induced XMCD changes sign upon reversing the magnetic field, reflecting its odd parity with respect to the external magnetic field.
Figure~3(e) presents the calculated XMLD spectra, which is the XAS difference between $E||[110]$ and $E||[\bar{1}10]$.
In the parent tetragonal $D_{4h}$ symmetry, the external magnetic field applied along the $z$ axis, $H_z$, transforms as the $A_{2g}$ irreducible representation.
Its coupling to the $B_{1g}$ magnetic octupole $M_{xyz}$ gives rise to a $B_{2g}$-symmetric electric-quadrupole ($Q_{xy}$) response according to
$B_{1g}\otimes A_{2g}=B_{2g}$,
as shown in Eq.~(4).
The calculated polarization difference reflects this $B_{2g}$ symmetry.
The XMLD signal exhibits odd parity with respect to the magnetic field, indicating that it is sensitive to the sign of the N\'eel vector.
The magnetic-field dependence of the XMLD intensity at $E = 638$~eV is summarized in Fig.~3(f).
The XMLD signal exhibits a leading-order linear dependence on the applied magnetic field and reverses sign between the two N\'eel-vector orientations ($\mathbf{N}\parallel[001]$ and $[00\bar{1}]$), indicating a symmetry-allowed linear coupling between the field and the spinful magnetic octupole order parameter.

Figure~3(g) shows the XMCD spectra under uniaxial pressure applied along the in-plane direction ($P||[110]$).
Although $4'/mm'm$ symmetry forbids magnetic dipolar responses,
an in-plane uniaxial strain $B_{2g}$ component, $\varepsilon_{xy}$, which breaks the fourfold rotational symmetry, $D_{4h}\rightarrow D_{2h}^{[110]}$.
The magnetic point group is reduced to an orthorhombic magnetic subgroup, $4'/mm'm\rightarrow m'm'm$.
The uniaxial stress $B_{2g}$ component couples to $B_{1g}$ magnetic octupole $M_{xyz}$ to form an $A_{2g}$ magnetic response: $B_{1g}\otimes B_{2g}=A_{2g}$, as shown in Eq.~(5).
Consequently, the symmetry breaking activates the
magnetic dipole response, thereby inducing a finite XMCD signal even in the absence of an external magnetic field.
The stress-induced XMCD reverses sign with the sign of the applied pressure, consistent with a linear piezomagnetic effect.
Figure~3(h) compares the XMCD spectra induced by the magnetic field ($H||c$) and by uniaxial pressure ($P||[110]$).
In MnF$_2$, applying uniaxial stress along the [110] direction lifts the degeneracy between sites 1 and 2 by generating an energy difference between them.
As a result, the oppositely oriented spins at the two sites respond at different energies in the spectrum, leading to a finite XMCD signal.
In contrast to MnTe, the magnetic dipole moments induced by the magnetic field and uniaxial pressure in MnF$_2$ are mainly from the conventional spin contribution.
Both perturbations couple to identical spin channels; thus, the resulting XMCD spectral line shapes are similar.
Figure~3(i) presents the uniaxial-pressure dependence of the XMCD intensity at $E = 638$~eV for the two N\'eel-vector orientations.
The XMCD response exhibits a leading-order linear dependence on pressure and reverses sign between opposite antiferromagnetic domains.
This behavior demonstrates that strain provides a direct handle to control and detect the spinful magnetic octupole order in MnF$_2$ through the piezomagnetic effect.

\begin{figure*}[t]
    \centering
    \includegraphics[width=0.9\textwidth]{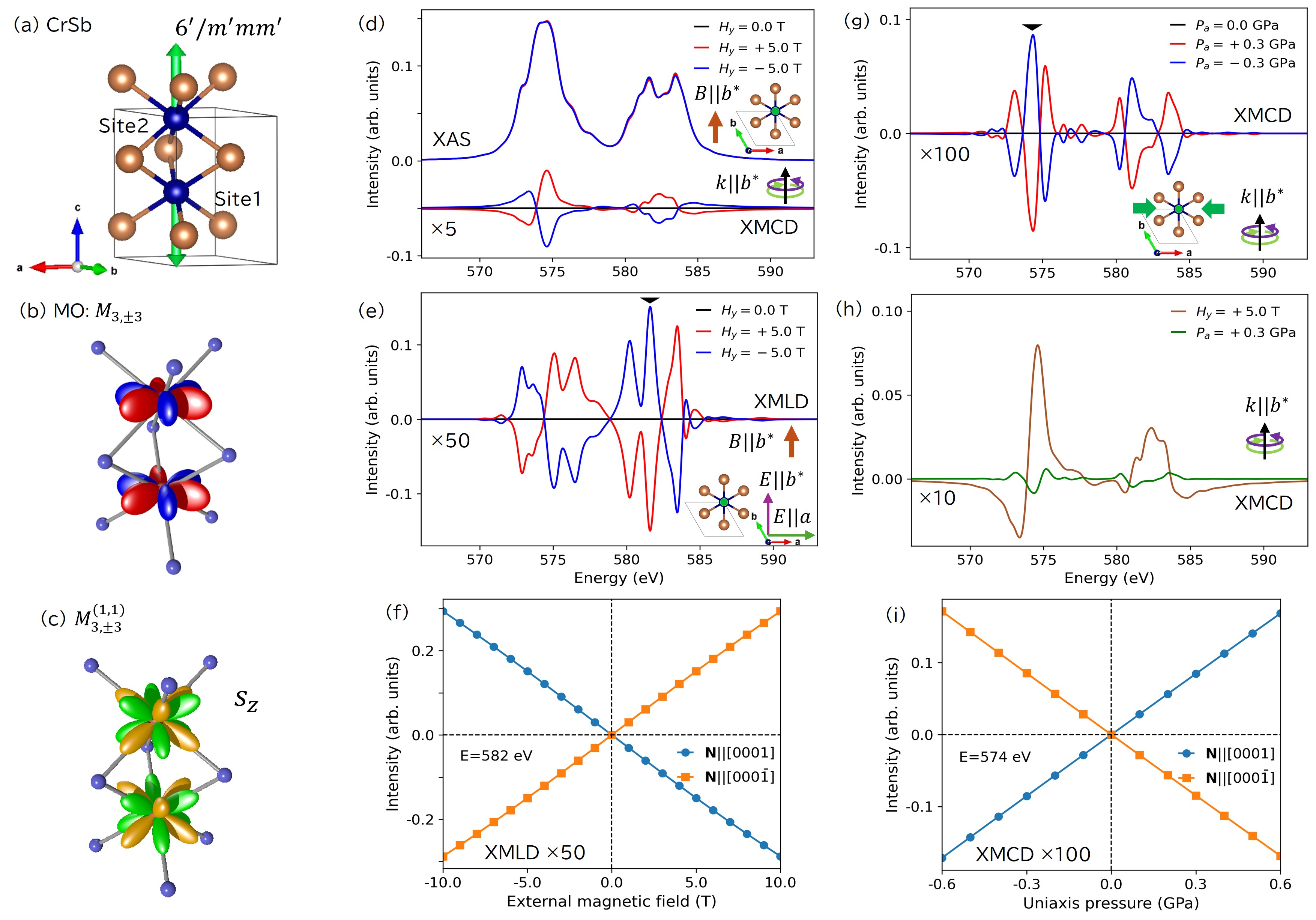}
    \caption{
    Magnetic multipole structure and calculated x-ray absorption spectra in CrSb.
    (a) Crystal and magnetic structure of CrSb with the magnetic point group $6'/m'mm'$.
    This symmetry permits
    (b) the $(x^3-3xy^2)$-like magnetic octupole moments ($M_{3,\pm3}$) and  (c) their spinful counterpart of the $z(x^3-3xy^2)$-type hexadecapole
    associated with the $s_z$ spin component, $M_{3,\pm3}^{(1,1)}$, where $x \parallel a$, $y \parallel b^{*}$, and $z \parallel c$.
    (d) X-ray absorption spectrum (XAS) and x-ray magnetic circular dichroism (XMCD) spectra around the Cr $L$ edge for magnetic fields applied parallel and antiparallel to the $b^{*}$ direction.
    (e) X-ray magnetic linear dichroism (XMLD) spectra measured between orthogonal linear polarizations ($E\parallel a$ and $E\parallel b^{*}$) under opposite magnetic fields.
    The XMLD signal reverses its sign upon inversion of the magnetic multipole domain, demonstrating its odd response with respect to the external magnetic field.
    (f) Magnetic-field dependence of the XMLD intensity at $E = 582$~eV.
    The signal exhibits a linear dependence on the applied magnetic field and changes sign depending on the N\'eel-vector orientation ($\mathbf{N}\parallel[0001]$ and $[000\bar{1}]$).
    (g) XMCD ($\mathbf{k}\parallel b^*$) spectra under uniaxial pressure ($P_{a}=\pm0.3$~GPa).
    The stress-induced XMCD emerges through piezomagnetic symmetry breaking even in the absence of net magnetization.
    (h) Comparison between magnetic-field-induced and uniaxial-pressure-induced XMCD spectra.
    (i) Uniaxial-pressure dependence of the XMCD intensity at $E = 574$~eV.
    A Lorentzian broadening of 0.42 eV (FWHM) was used for CrSb. The sign and qualitative line shape of the XMCD/XMLD spectra are robust against reasonable variations of the broadening parameter ($\pm 0.1$ eV).
    }
    \label{fig:CrSb}
\end{figure*}

\subsection{$g$-wave altermagnet}
CrSb shares the same NiAs-type crystal structure as $\alpha$-MnTe, belonging to the hexagonal space group $P6_3/mmc$ (No.~194) \cite{Willis1953_CrSb, Nagasaki1969_CrSb_HP}.
Below its N\'eel temperature ($T_N \sim 710$~K \cite{NeelTmpCrSb}), CrSb exhibits a collinear antiferromagnetic order in which the Cr spins are aligned along the $c$ axis [Fig.~4(a)].
This magnetic configuration gives rise to the magnetic point group $6'/m'mm'$, reflecting the rotation of the N\'eel vector from the $ab$ plane in MnTe to the $c$ axis in CrSb.
Under the $6'/m'mm'$ symmetry, rank-1 magnetic dipole responses along the $c$ axis are strictly forbidden, ensuring zero net magnetization and no intrinsic dipolar XMCD at zero field.
However, higher-rank magnetic multipoles are symmetry allowed.
In particular, the rank-3 magnetic octupole components $M_{3,\pm3}$ are permitted [Fig.~4(b)].
The microscopic origin of these octupolar moments is described by the spinful operator $M_{3,\pm3}^{(1,1)}$ \cite{Yatsushiro2021PRB}, which combines the $s_z$ spin component with a $z(x^3-3xy^2)$-type angular distribution [Fig.~4(c)].
The ferroic magnetic octupole order $M_{3a}$ transforms as $B_{1g}$ of $D_{6h}$.

Figure~4(d) shows the calculated XAS and XMCD spectra around the Cr $L$ edge for incident x-rays with $\mathbf{k}\parallel b^*$.
Consistent with symmetry, no intrinsic dipolar XMCD appears at zero magnetic field.
When a magnetic field is applied perpendicular to the $c$ axis, $e.g.$, along the $b^*$ direction, XMCD signals emerge due to the canted spin components.
The applied magnetic field gives rise to a reduction of the sixfold rotational symmetry, resulting in a finite XMLD (XAS difference between $E||x~[11\bar{2}0]$ and $E||y~[1\bar{1}00]$) signal, as shown in Fig.~4(e).
This behavior indicates that the electric quadrupole response is linearly coupled to the magnetic octupole moment through the external field, as depicted by Eq.~(4),
schematically expressed as a product of the magnetic octupole and the magnetic field.
In the parent hexagonal symmetry, the in-plane magnetic field ($H_x$, $H_y$) transforms as the $E_{1g}$ irreducible representation. Its coupling to the $B_{1g}$ magnetic octupole $M_{3a}$ gives rise to an $E_{2g}$ electric-quadrupole ($Q_{xy}$, $Q_{x^2-y^2}$) response according to
$B_{1g}\otimes E_{1g}=E_{2g}$,
as shown in Eq.~(4).
The calculated polarization difference reflects this $Q_{x^2-y^2}$ symmetry.
Microscopically, the applied magnetic field modifies the relative stability of electronic states determined by the interplay between ligand-field anisotropy and spin orientation.
As a result, field-induced redistribution of orbital occupations occurs, leading to a reversal of the XMLD sign.
The underlying anisotropic electronic structure is closely related to that discussed for $\alpha$-MnTe, $i.e.$,
where the in-plane spin component modulates the spatial distribution of $e_g^\pi$ orbitals.
Figure~4(f) presents the magnetic-field dependence of the XMLD intensity at $E = 582$~eV.
The signal exhibits a leading-order linear dependence on the magnetic field and reverses sign between the two N\'eel-vector orientations ($\mathbf{N}\parallel[0001]$ and $[000\bar{1}]$), demonstrating linear coupling between the magnetic field and the octupole order parameter.

Figure~4(g) shows the XMCD spectra measured with $\mathbf{k}\parallel b^*$ under uniaxial stress applied along the crystallographic $a$ axis in the absence of an external magnetic field.
The unstrained antiferromagnetic state of CrSb belongs to the magnetic point group $6'/m'mm'$, which forbids rank-1 magnetic-dipole responses.
The uniaxial stress along the $a$ axis contains, in addition to a symmetry-preserving $A_{1g}$ component, an $E_{2g}$ strain component,
$\sigma_{E_{2g}}$.
Because $B_{1g}\otimes E_{2g}=E_{1g}$,
the composite quantity $M_{3a}\sigma_{E_{2g}}$ transforms as an in-plane magnetic dipole $(M_x,M_y)$.
For the present stress geometry, the induced component is $M_y\parallel b^*$, and the magnetic point group is reduced according to $6'/m'mm'\rightarrow m'_{x}m_{y}m'_{z}$,
where the orthorhombic axes are defined as $x\parallel a$, $y\parallel b^*$, and $z\parallel c$.
Consequently, an $E_{1g}$-symmetric rank-1 magnetic response is activated along the $b^*$ direction,
giving rise to finite XMCD for $\mathbf{k}\parallel b^*$ even in the absence of an external magnetic field.
The stress-induced XMCD reverses sign upon reversal of either the $E_{2g}$ strain or the magnetic-octupole domain, consistent with the linear piezomagnetic coupling to the preexisting $M_{3a}$ order.
Figure~4(h) compares the XMCD spectra induced by the magnetic field and uniaxial pressure.
Since their microscopic origins differ, the spin component for the magnetic field and the AMD component from the magnetic octupole for strain, the spectral line shapes are distinct.
Finally, Fig.~4(i) shows the uniaxial-pressure dependence of the XMCD intensity at $E = 574$~eV for the two opposite N\'eel-vector orientations.
The leading-order linear and sign-reversing behavior suggests that strain can induce and probe the magnetic octupole order in CrSb, reflecting a symmetry-allowed coupling between magnetic octupole order and electric quadrupole responses in this $g$-wave altermagnet.

\section{Conclusions}
We have presented a symmetry-based spectroscopic framework for piezomagnetic and inverse piezomagnetic effects in altermagnets hosting ferroic magnetic dipole and octupole orders.
By formulating the problem in the complete multipole basis, magnetic point groups can be systematically classified, and the allowed third-rank response tensors that connect magnetization, stress, magnetic field, and electric quadrupoles can be identified in a unified manner.
This multipolar language provides a transparent criterion to distinguish dipolar, quadrupolar, and octupolar responses, even in systems with vanishing net magnetization.
These couplings give rise to characteristic anisotropies and field/strain responses that serve as clear signatures of $d$- and $g$-wave altermagnets, also providing experimentally accessible fingerprints of hidden multipolar order.
By combining symmetry analysis with element-specific XMCD and XMLD spectroscopy, we have presented numerical examples and qualitative spectral fingerprints of piezomagnetic and inverse piezomagnetic symmetry breaking. While the calculated spectra provide symmetry-consistent indicators of the underlying multipolar order, they do not constitute a general method for uniquely assigning every observed XMCD/XMLD spectral feature to a specific microscopic multipolar origin.
Since strain and magnetic field couple to the same underlying multipolar channels, magnetoelastic engineering offers a powerful route to manipulate antiferromagnetic domain states without relying on net magnetization.
This framework thus opens a pathway toward strain-controllable spin functionalities and multipole-based device concepts that extend beyond conventional dipolar spintronics, highlighting altermagnets as a fertile platform for next-generation symmetry-driven spin phenomena.

\begin{acknowledgments}
The authors thank M. Mizumaki, T. Uozumi, and T. Arima for constructive discussions.
This project is partly supported by the Japan Society for the Promotion of Science (JSPS) KAKENHI (Grants No.~JP19H04399, No.~JP24K03205, No.~JP24H01685, and No.~JP24K17603).
This work was supported by the MEXT Quantum Leap Flagship Program (MEXT Q-LEAP) under Grant No.~JPMXS0118068681.
This work is also partially supported by CREST (Grant No.~JPMJCR2435), Japan Science and Technology Agency (JST).
\end{acknowledgments}

\section*{Data availability}
There are no publicly available research data or software supporting this manuscript. Requests for further information or data should be sent to the authors.

\appendix*
\section{Dependence of external field and stress responses on the effective exchange field in altermagnets}
\label{sec:appendix_exchange_dependence}

In this Appendix, we systematically investigate the sensitivity of the magnetoelastic-induced x-ray spectroscopic responses to the magnitude of the effective exchange field ($B_{\text{int}}$) in altermagnetic materials. Specifically, we evaluate the spectra for $\text{MnTe}$, $\text{MnF}_2$, and $\text{CrSb}$ across a wide range of exchange fields: $B_{\text{int}} = 0.001$, $0.002$, $0.005$ (the reference value used in the main text), and $0.01\text{ eV}$.
Here, $B_{\text{int}}$ enters the mean-field Hamiltonian in the same form as the Zeeman term, $2\boldsymbol{B}_{\text{int}}^{(i)}\cdot\boldsymbol{s}$, as in our previous studies~\cite{sasabe2023ferroic, Sasabe2025PRB}.
Throughout these calculations, the external magnetic field was set to $+5\text{ T}$ for the field-induced spectroscopic response, and the applied uniaxial stress was set to $+0.3\text{ GPa}$ for the stress-induced response. The geometries of external fields, applied stress, and x-ray incident directions for each material are identical to the experimental configurations described in the main text.

\subsection{MnTe}
\label{subsec:appendix_MnTe}

\begin{figure}[t]
    \centering
    \includegraphics[width=0.8\linewidth]{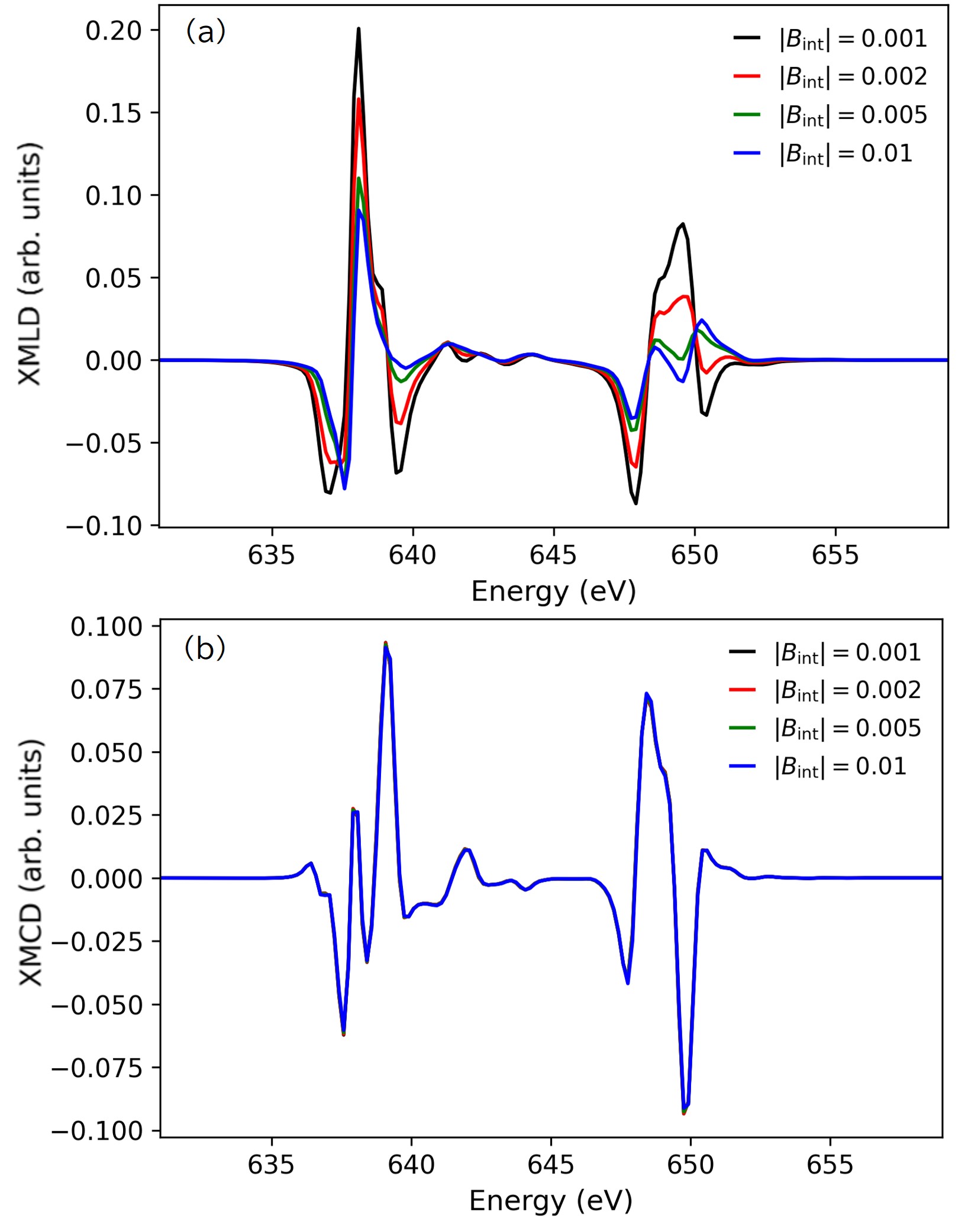}
    \caption{
        Dependence of the magnetoelastic responses in MnTe on the effective exchange field ($B_{\text{int}}$).
        (a) Field-induced x-ray magnetic linear dichroism (XMLD) spectra under a perpendicular external magnetic field of $+5\text{ T}$.
        Due to the geometry where the applied magnetic field is orthogonal to the N\'eel vector, the XMLD spectral intensity is enhanced as $B_{\text{int}}$ decreases.
        (b) Stress-induced x-ray magnetic circular dichroism (XMCD) spectra under a uniaxial stress of $+0.3\text{ GPa}$.
        The XMCD response is nearly independent of $B_{\text{int}}$.
        The external field and stress geometries are identical to those in the main text.
        A Lorentzian broadening of 0.42 eV (FWHM) was used.
    }
    \label{fig:appendix_MnTe}
\end{figure}

Figure~\ref{fig:appendix_MnTe} shows the calculated spectra of (a) the field-induced XMLD under a perpendicular external magnetic field of $+5\text{ T}$ and (b) the stress-induced XMCD under a uniaxial stress of $+0.3\text{ GPa}$ for MnTe, as a function of the effective exchange field ($B_{\text{int}}$).
In MnTe, the N\'eel vector and the applied magnetic field are orthogonal, a geometry where the external field can readily induce spin canting.

As shown in Fig.~\ref{fig:appendix_MnTe}(a), the intensity of the field-induced XMLD is significantly enhanced as $B_{\text{int}}$ decreases.
Physically, a smaller internal exchange coupling corresponds to a lower spin stiffness of the altermagnetic order.
This reduced stiffness makes the spins more susceptible to canting toward the external field direction.
Since XMLD directly probes the resulting orbital anisotropy (even-rank electric quadrupole moments), this facilitated spin canting directly manifests as a prominent enhancement of the XMLD spectrum.
Qualitatively, this behavior mimics the finite-temperature effect in real materials, where thermal fluctuations soften the exchange stiffness and promote field-induced spin reorientation.

In sharp contrast, the stress-induced XMCD spectrum in Fig.~\ref{fig:appendix_MnTe}(b) exhibits negligible dependence on $B_{\text{int}}$.
The line shape and intensity remain almost perfectly constant across the entire range from $0.001$ to $0.01\text{ eV}$.
This robustness is reasonable because, for all values of $B_{\text{int}}$ considered, the sublattice spins remain fully polarized along their exchange fields. The stress-induced XMCD arises from the stress-induced change in the crystal-field environment rather than from spin canting, and is therefore insensitive to $B_{\text{int}}$.

This striking contrast between the highly sensitive XMLD and the robust XMCD highlights that our theoretical framework correctly captures the distinct roles of different multipolar orders and their responses to external stimuli.
The qualitative features of the magnetoelastic responses are universally governed by symmetry, independent of the specific magnitude of the exchange interaction.

\subsection{MnF$_2$}

\label{subsec:appendix_MnF2}

\begin{figure}[t]
    \centering
    \includegraphics[width=0.8\linewidth]{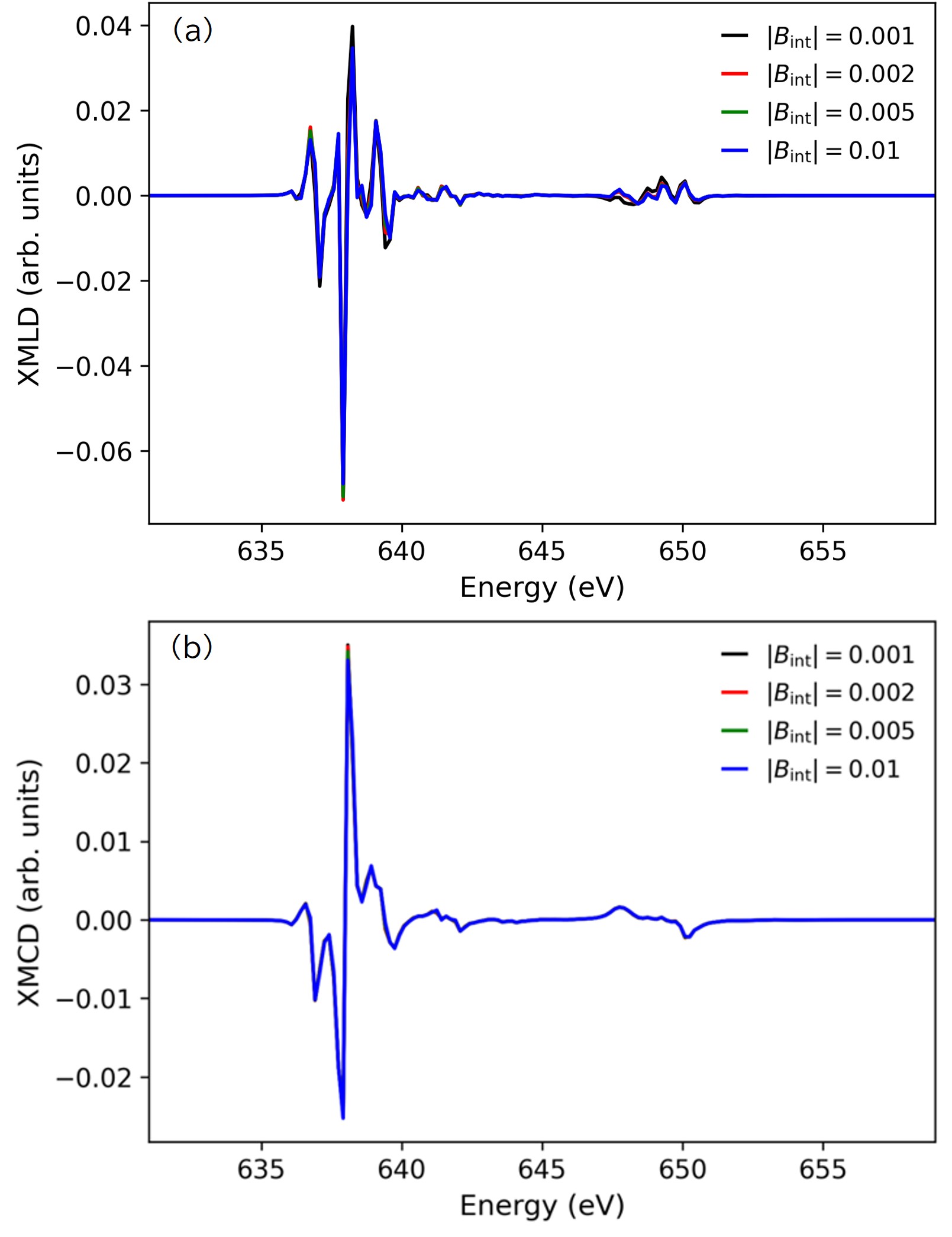}
    \caption{
        Dependence of the magnetoelastic responses in MnF$_2$ on the effective exchange field ($B_{\text{int}}$).
        (a) Field-induced x-ray magnetic linear dichroism (XMLD) spectra under a parallel external magnetic field of $+5\text{ T}$.
        Due to the geometry where the applied magnetic field is parallel to the N\'eel vector, spin canting is suppressed, resulting in a spectrum that is independent of $B_{\text{int}}$.
        (b) Stress-induced x-ray magnetic circular dichroism (XMCD) spectra under a uniaxial stress of $+0.3\text{ GPa}$.
        The XMCD response is nearly independent of $B_{\text{int}}$.
        The external field and stress geometries are identical to those in the main text.
        A Lorentzian broadening of 0.17 eV (FWHM) was used.
    }
    \label{fig:appendix_MnF2}
\end{figure}

For the rutile-type altermagnet $\text{MnF}_2$, we performed the same sensitivity analysis, with the results shown in Fig.~\ref{fig:appendix_MnF2}.
In stark contrast to MnTe, the magnetic configuration of $\text{MnF}_2$ features a N\'eel vector aligned parallel to the direction of the applied external magnetic field.
In this geometry, the torque exerted by the external field on the antiferromagnetic sublattices is strongly constrained by the geometric alignment, thereby suppressing any significant spin canting.

As a direct result of this geometrically constrained configuration, the field-induced XMLD spectrum, shown in Fig.~\ref{fig:appendix_MnF2}(a), exhibits a negligible dependence on the effective exchange field $B_{\text{int}}$.
The spectra remain almost identical across the entire tested range from $0.001$ to $0.01\text{ eV}$.

On the other hand, and consistent with the case of MnTe, the stress-induced XMCD spectrum under $+0.3\text{ GPa}$, shown in Fig.~\ref{fig:appendix_MnF2}(b), is remarkably robust against variations in $B_{\text{int}}$.
This again demonstrates that the spectroscopic signature of the stress-induced odd-rank magnetic multipoles is fundamentally protected by symmetry.

The contrasting behavior between MnTe and $\text{MnF}_2$ powerfully corroborates our central conclusion: the nature of the magnetoelastic response (i.e., whether it is sensitive to spin stiffness or not) is dictated by the geometric alignment of the magnetic order with respect to the external field, while the qualitative spectral features are universally governed by the underlying symmetry.

\subsection{CrSb}

\label{subsec:appendix_CrSb}

\begin{figure}[t]
    \centering
    \includegraphics[width=0.8\linewidth]{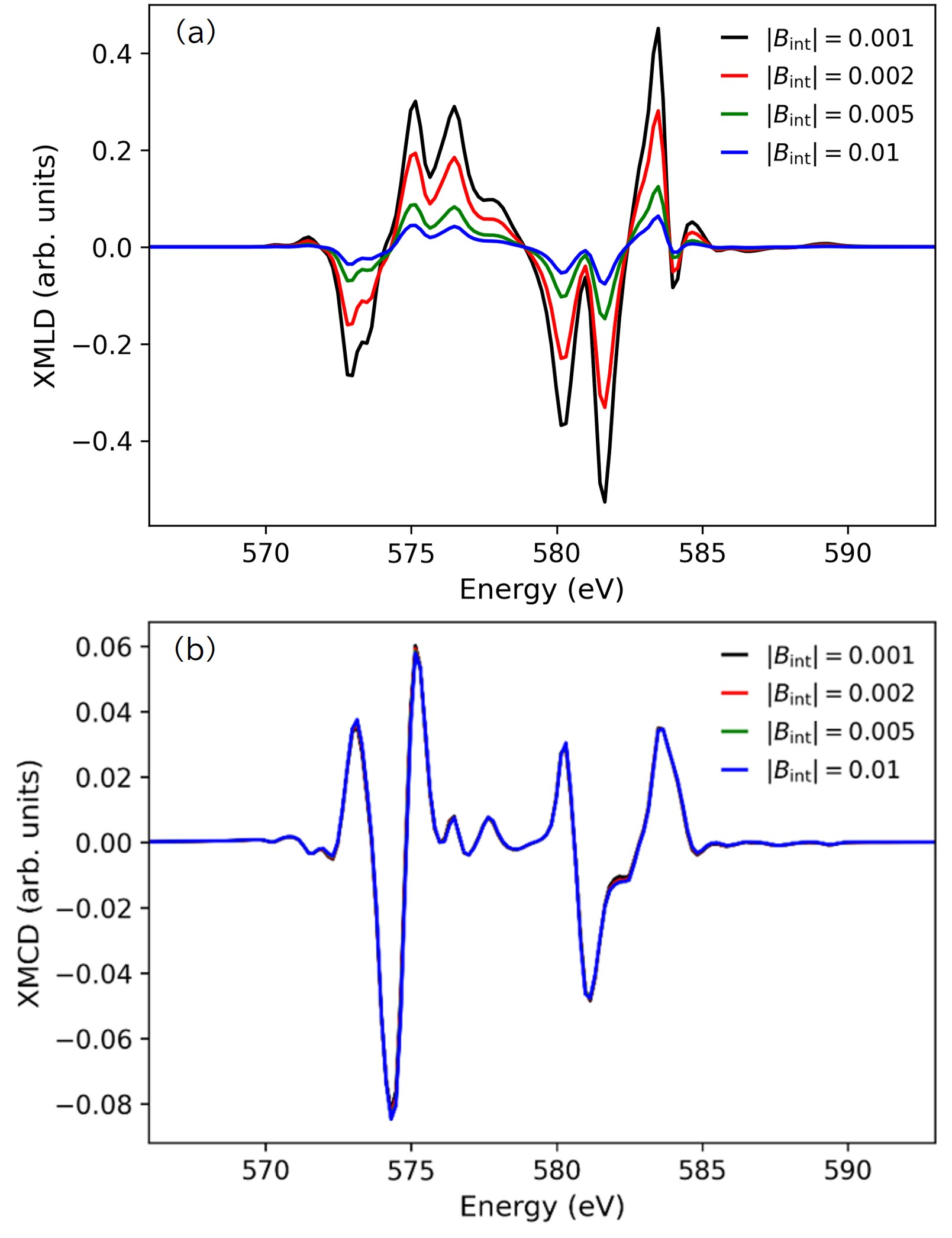}
    \caption{
        Dependence of the magnetoelastic responses in CrSb on the effective exchange field ($B_{\text{int}}$).
        (a) Field-induced x-ray magnetic linear dichroism (XMLD) spectra under a perpendicular external magnetic field of $+5\text{ T}$.
        Due to the geometry where the applied magnetic field is orthogonal to the N\'eel vector, the XMLD spectral intensity is enhanced as $B_{\text{int}}$ decreases.
        (b) Stress-induced x-ray magnetic circular dichroism (XMCD) spectra under a uniaxial stress of $+0.3\text{ GPa}$.
        The XMCD response is nearly independent of $B_{\text{int}}$.
        The external field and stress geometries are identical to those in the main text.
        A Lorentzian broadening of 0.42 eV (FWHM) was used.
    }
    \label{fig:appendix_CrSb}
\end{figure}

Finally, we present the sensitivity analysis for the metallic altermagnet $\text{CrSb}$ in Fig.~\ref{fig:appendix_CrSb}.
Similar to MnTe, the magnetic configuration of $\text{CrSb}$ features a N\'eel vector oriented orthogonally to the applied external magnetic field.
This geometry allows the external field to exert a torque on the magnetic sublattices, thereby inducing spin canting.

As expected from this geometry and consistent with the case of MnTe, the field-induced XMLD spectrum, shown in Fig.~\ref{fig:appendix_CrSb}(a), is highly sensitive to the magnitude of $B_{\text{int}}$.
The spectral intensity is significantly amplified in the weak-exchange limit (e.g., $B_{\text{int}} = 0.001\text{ eV}$), where the reduced spin stiffness facilitates a larger canting angle.
Conversely, the stress-induced XMCD spectrum, shown in Fig.~\ref{fig:appendix_CrSb}(b), once again demonstrates remarkable robustness, with the spectra remaining virtually unchanged across the entire range of $B_{\text{int}}$.

The consistent trend observed across all three distinct altermagnetic compounds,
MnTe, MnF$_2$, and CrSb,
provides compelling and conclusive evidence that the spectroscopic signatures of magnetoelasticity are universally dictated by the underlying symmetries and geometric configurations.
The qualitative nature of these responses is not an artifact of a specific exchange interaction strength but is fundamentally protected by symmetry, solidifying the central conclusions of our work.

\input{bib_ssb.bbl}
\end{document}

%% file: bib_ssb.bbl
\providecommand*{\bibhyphen}{-}